# Modeling of intermediate structures and chain conformation in silica-latex nanocomposites observed by SANS during annealing

Anne-Caroline Genix[1,2], Mouna Tatou[1,2,4], Ainara Imaz[3], Jacqueline Forcada[3],

Ralf Schweins[4], Isabelle Grillo[4], Julian Oberdisse[1,2,5]

[1] *Université Montpellier 2, Laboratoire Charles Coulomb UMR 5221, F-34095 Montpellier, France*

[2] *CNRS, Laboratoire Charles Coulomb UMR 5221, F-34095 Montpellier, France*

[3] *Grupo de Ingeniería Química, Facultad de Ciencias Químicas, The University of the Basque Country, 20080 San Sebastián, Spain*

[4] *Institut Laue-Langevin, F-38042 Grenoble, France*

[5] *Laboratoire Léon Brillouin, UMR 12 CEA/CNRS, CEA Saclay, F-91191 Gif sur Yvette, France*

\* author for correspondence : acgenix@univ-montp2.fr

February 2012

**Abstract:**

The evolution of the polymer structure during nanocomposite formation and annealing of silica-latex nanocomposites is studied using contrast-variation small angle neutron scattering. The experimental system is made of silica nanoparticles ($R_{si} \approx 8$ nm) and a mixture of purpose-synthesized hydrogenated and deuterated nanolatex ($R_{latex} \approx 12.5$ nm). The progressive disappearance of the latex beads by chain interdiffusion and release in the nanocomposites is analyzed quantitatively with a model for the scattered intensity of hairy latex beads and an RPA description of the free chains. In silica-free matrices and nanocomposites of low silica content (7%v), the annealing procedure over weeks at up to $T_g + 85$ K results in a molecular dispersion of chains, the radius of gyration of which is reported. At higher silica content (20%v), chain interdiffusion seems to be slowed down on time-scales of weeks, reaching a molecular dispersion only at the strongest annealing. Chain radii of gyration are found to be unaffected by the presence of the silica filler.

**Figures: 7**
**Tables: 5**

## I. Introduction

Dispersing hard filler nanoparticles in a soft polymeric matrix creates a new material, nanocomposites. Their properties may be tuned, e.g., by modifying the quality of the dispersion, or of the interfacial interactions.[1-5] Understanding how properties of nanocomposites emerge necessitates knowledge of polymer structure and dynamics, as well as of nanoparticle dispersion. General fundamental results are rare in this field, and many model systems have been investigated in the past. Focusing on chain conformation, one finds various systems, like silica particles in poly(dimethylsiloxane) (PDMS)[6], poly(styrene) (PS)[7, 8], poly(isoprene)[9, 10] or poly(ethylene-propylene) (PEP)[11], and crosslinked PS beads in PS[12]. The subject has also been addressed by computer simulations[13-16] and molecular theories[17, 18]. The conclusion on the evolution of the chain radius of gyration with the amount of hard filler in most of these studies remains incomplete at the moment, among others due to difficulties in measuring a chain structure which is not perturbed by the filler presence. It is not clear to date, if this is a technical problem of, e.g., impossible matching of heterogeneous nanoparticles, or if the chain structure is truly perturbed by obstacles.

Polymer dynamics in a hard, nanostructured filler environment is one of the intriguing problems in nanocomposites. Access to this information has been gained using fluorescence resonance energy transfer[19], quasielastic neutron scattering[20-23], and NMR techniques[24-29]. Reorganization of polymer molecules in melts, e.g., by following the radius of gyration of polymer chains during annealing by scattering, gives information on larger scales[30-33]. Inspired by this method, we have recently chosen a similar approach[34], which is also used here.

A large body of experimental research has been dedicated to nanoparticle dispersions in nanocomposites[35]. A particularly interesting system is based on latex film formation. The process of formation of pure latex films has attracted considerable attention over the past two decades[36-44]. Latex beads are brought into contact by evaporation of the aqueous solvent above the minimal film formation temperature. Depending on the thermal history of the samples and the architecture of the beads, different structures may be generated. For large beads with a clearly defined shell, e.g., beads may keep their globular shape, and the shells connect into a network. In other cases, all chains may interdiffuse until the initial organization in the form of beads has completely disappeared, and the final film forms a molecularly dispersed melt of polymer chains.

Latex film formation has been used as a method to incorporate silica nanoparticles into a polymer melt, thereby forming silica-latex nanocomposites [19]. One of the advantages of the silica-latex nanocomposite system is that the silica structure in the final nanocomposites can be controlled by the physico-chemical properties of the precursor solution [34, 45, 46]. The samples chosen in the present article have a similar silica structure at different filler volume fractions, due to a simultaneous change of solution pH and silica concentration. Another advantage is that one may mix hydrogenated (H) and deuterated (D) latex nanoparticles, and thereby create a scattering contrast for the polymer, allowing its structural analysis by neutron scattering. Under particular conditions, called zero-average contrast (ZAC) [47, 48], the influence of the silica-filler may be strongly reduced. In a recent paper [34], we have studied the rheology and silica structure of such nanocomposites. We have introduced D-latex particles, and the scattering was shown to be dominated by the polymer structure. During annealing, H and D-latices were found to demix due to incompatibility in the physico-chemistry of the bead stabilization. The analysis of the demixing kinetics gave information on the bead dynamics in presence of the hard filler phase: low silica volume fractions (5%) had little impact on the demixing kinetics, whereas high volume fractions (15%) were sufficient to block the demixing on the time scale of observation.

In this article, the structural evolution during annealing of films made of H- and D-nanolatex beads, which are *compatible* due to the use of the same stabilization layer, is discussed. Following the ZAC-concept, it will be shown that the silica signal can be made negligible in nanocomposites. Small angle neutron scattering (SANS) then gives access to polymer structure on the nanometer scale, and to slow dynamics over time scale of weeks. This study thus opens a route to characterizing the chain structure in nanocomposites, as a function of the filler quantity and dispersion[49]. Materials, synthesis of H- and D-latices, and experimental methods are presented in section II. Experimental results are presented in section III, starting with the phenomenology of the structural evolution during annealing (III.1). This is followed by a quantitative modeling based on a combination of the Pedersen model of hairy beads, and of the classical RPA equation for free chains. The model is applied to the experimental intensities of H-D structures both in silica-free matrices (III.2), and in nanocomposites (III.3).

In section III.4, the impact of the silica-content on chain dynamics is discussed, before concluding with perspectives in section IV.

**II. Materials and methods**

**Silica nanoparticles:** Bindzil silica nanoparticles delivered in high pH charge-stabilized aqueous suspensions (30%wt, pH 9 - 9.5) were a gift from Akzo Nobel. We have checked by SANS that they are individually dispersed, and their dimensions are described by a log-normal size distribution ($R_{LN}$ = 78.5 Å and $\sigma$ = 18%) leading to an average volume of $V_{si}$ = 2.34 $10^6$ Å$^3$ and a volume-average radius of $R_{si}$ = 82 Å. Contrast variation was employed to determine the scattering length density ($\rho_{si}$ = 3.6 $10^{10}$ cm$^{-2}$).[34] The hydrodynamic radius determined by dynamic light scattering is 120 Å.

**Synthesis of hydrogenated and deuterated latex nanoparticles:** Hydrogenated polymer nanoparticles (H-latex) were purpose-synthesized in San Sebastian using semicontinuous emulsion copolymerization of methyl methacrylate (MMA) and butyl acrylate (BuA). The surfactant used for stabilisation was sodium dodecyl sulfate (SDS, Merck) (6%wt with respect to the total monomer mass). The synthesis process has been described elsewhere [34, 50]. Deuterated polymer nanoparticles (D-latex) were synthesized following the same protocol, with perdeuterated MMA (containing 8 D), and BuA containing 9 D and 3 H.

**Characterization of hydrogenated and deuterated latex nanoparticles:** The composition of both H- and D-latex beads has been checked. The molar fractions of MMA (72 ± 1%) and BuA (28 ± 1%) have been measured by $^1$H-NMR for protonated batches and by $^{13}$C-NMR for deuterated ones, both in CDCl$_3$. The glass-transition temperature was found to be 35°C on average by differential scanning calorimetry. The average chain mass has been obtained by size exclusion chromatography (SEC) using hydrogenated PMMA-standards in THF, and is given, together with their polydispersity index, in **Table 1**. Due to limited batch size with deuterated material, four different batches were synthesized, called batches A to D. The macroscopic density was found to be $d_H$ = 1.16 ± 0.02 g.cm$^{-3}$ for H (resp. $d_D$ = 1.23 ± 0.02 g.cm$^{-3}$ for D) at room temperature, leading to monomer (repeat unit) volumes of $V_H$ = 1.54 $10^{-23}$ cm$^3$ (resp. $V_D$ = 1.57 $10^{-23}$ cm$^3$). The scattering length density was determined by independent contrast variation experiments (see appendix): $\rho_H$ = 0.94 $10^{10}$ cm$^{-2}$ (resp. $\rho_D$ = 6.4 $10^{10}$ cm$^{-2}$), in agreement with the macroscopic density and the composition. The index-match point of the silica nanoparticles is thereby found to be at a matrix volume fraction in hydrogenated polymer of $\Phi_H = 51\%$ ($\Phi_H + \Phi_D = 100\%$). The scattering length density of the H-D-matrix is then given by $\rho_{HD} = \Phi_H \rho_H + \Phi_D \rho_D$. The nomenclature of the samples goes as follows: the letter corresponds to the batch, $\Phi_H$ is given in the index, and where appropriate, the silica volume fraction in parenthesis (e.g., $A_{53}(7\%)$ for a sample made of batch A, having 53% H, and 7% silica).

| Name | $M_w(H)$ /PI (g/mol) | $M_w(D)$ /PI (g/mol) | pH | $\Phi_H$ | $\chi_s$ |
|---|---|---|---|---|---|
| $A_{53}$ | 279 100 /2.7 | 217 900 /2.5 | 9 | 0.53 | $9.3\ 10^{-4}$ |
| $A_{62}$ | 279 100 /2.7 | 217 900 /2.5 | 9 | 0.62 | $10.1\ 10^{-4}$ |
| $B_{62}$ | 258 200 /2.5 | 242 100 /2.3 | 4 | 0.62 | $9.7\ 10^{-4}$ |
| $C_{44}$ | 305 500 /3.3 | 240 200 /2.6 | 7 | 0.44 | $8.3\ 10^{-4}$ |
| $D_{51}$ | 345 100 /3.4 | 241 000 /1.9 | 5 | 0.51 | $8.0\ 10^{-4}$ |

**Table 1:** Characteristics of the different polymer matrices. Chain masses of H- and D-latex were obtained by SEC, pH refers to the precursor solution, $\Phi_H$ to the volume fraction of H-latex in the matrix, and $\chi_s$ is the theoretical monomer interaction parameter on the spinodal curve calculated according to **eq.(7)**.

The size distribution of the various batches of latex beads (H and D) has been characterized by SANS in dilute suspension. The synthesis protocol gave H- and D-latex particles of typical radius $R_{Guinier} = R_g*\sqrt{(3/5)} = 125$ Å. This corresponds to the average mass or dry volume measured by $I(q\rightarrow 0)$, which is $8.2\ 10^6$ Å$^3$. The average mass of the H- and D-chains (280 and 230 kg/mol, resp.) of the matrices used for the annealed samples was taken to estimate the typical number of chains per latex bead: $N_0 = 23$.

**Nanocomposite formulation and film formation:** Bubble-free silica-latex films were formed by slow drying in teflon moulds at 65°C during three days, after deionisation and degassing of solutions, and immediate pH adjustment to the desired value using NaOH. Thermo-gravimetric analysis was used to determine the silica volume fraction in the samples. D-containing samples for scattering discussed here were thinner (0.2 - 0.3 mm) than the H-latex films used to determine the silica structure by SANS (cf. appendix), to avoid multiple scattering. Annealing was performed at several temperatures well above $T_g$ (100°C to 120°C; one test at 150°C), over periods of one or two weeks. Cooling down samples rapidly below their glass-transition temperature $T_g$ then freezes the structures established at high temperature. Their characterization thus conveys information on polymer structure and interactions at the annealing temperature.

**Nanocomposite film structure determination:** The generic structure of silica nanoparticles dispersed in latex matrices has been studied previously [34]. For samples with silica, the filler structure has been checked by SANS in H-latex, and is reported in the appendix. For all other samples discussed in this article, the silica was matched by using appropriate mixtures of H- and D-latex. The evolution of the film structure from latex nanoparticles to molecularly dispersed chains was followed by SANS, as a function of thermal history and silica content. Small Angle Neutron Scattering was performed on beamlines D22 at Institut Laue-Langevin (ILL) (three configurations, defined by sample-to-detector distance D and incident neutron wavelength $\lambda$: D = 17 m; D = 8 m; D = 2 m, all $\lambda$ = 6 Å) and PACE at Laboratoire Léon Brillouin (LLB) (D = 4.5 m, $\lambda$ = 12 Å; D = 4.5 m, $\lambda$ = 6 Å; D = 1 m, $\lambda$ = 6 Å). Empty cell or empty beam subtraction, calibration by 1mm light water in Hellma cuvettes, and absolute determination of scattering cross-sections $I(q) = d\Sigma/d\Omega$ per unit sample volume in cm$^{-1}$ were performed using standard procedures[51], namely an incoming beam measurement for absolute units. Incoherent background was estimated using a far-point method, and checked by the known high-q scattering laws of polymeric interfaces and melts.

## III. Results

### III.1 Structural evolution during annealing of pure H-D-matrices

We start with the key observations by SANS of the progressive interdiffusion of H- and D-polymer chains in pure polymer films (i.e., silica-free matrices), before modeling them quantitatively in section III.2, and studying H-D-silica nanocomposites in section III.3. In the course of the annealing procedure, the initial film structure of possibly deformed latex-beads in close contact is replaced by molecularly dispersed chains. In a scattering experiment, the two limiting cases correspond to very different signatures in q-space. Both can be seen as space-filling objects. Given that the initial bead dispersion is governed by the structure of the (drying) suspension, and that H- and D-beads contain different isotopes but are otherwise identical, there should be no preferential correlation between H- and D-beads. In particular, contrarily to our previous study [34], both latex beads bear the same stabilizing layer. Then the theorem describing melt scattering can be applied to an ideal and incompressible mixture of H- and D-objects of normalized form factor P(q) [47, 48, 52]:

$$I(q) = \Delta\rho^2 \Phi_H \Phi_D v^* P(q) \qquad (1)$$

where $\Delta\rho$ is the scattering length density difference between H- and D-objects $\Delta\rho = \rho_H - \rho_D$, $\Phi_i$ their volume fraction ($\Phi_H + \Phi_D = 1$), v* their volume, and $P(q\rightarrow 0) = 1$. The theorem simply states that in the ideal case, no bead interaction parameter $\chi$ describing possible concentration fluctuations is needed, and only form factor scattering is observed. For globular particles, the typical scattering is thus the one of beads of high mass (expressed through the volume v*), i.e., high $I_o = I(q\rightarrow 0)$. A typical radius of gyration can be identified in the Guinier regime. A Porod law ~ $1/q^4$ in the high-q domain corresponds to its smooth interface. The other extreme – individual chains – have necessarily a lower $I_o$, because several chains make up one bead, and the ratio of $I_o$ can be used to extract the average number of chains per bead. Their high-q power law is also different, proportional to $1/q^2$, which corresponds to Gaussian chain statistics in a melt [53] in our q-range. The form factor of such Gaussian chains reads [54]:

$$P_{Debye}(q) = \frac{2}{q^4 R_g^4}\left(\exp(-q^2 R_g^2) - 1 + q^2 R_g^2\right) \qquad (2)$$

where $R_g$ is the radius of gyration of the macromolecules.

While the mixture of H- and D-latex beads in solution is likely to be ideal, because it is random in the latex suspension just before the gel point, the melts of H- and D-chains are usually not ideal: a monomeric (Flory-Huggins) interaction parameter $\chi$ between H and D monomers has to be introduced [55-57]. **Eq. 1** is then extended to the well-known random-phase approximation (RPA) [58]. In our notations, the scattering function for a mixture of H- and D-labelled chains of different masses and normalized chain form factor $P_H(q)$ ($P_D(q)$, respectively) is then given by

$$\frac{\Delta\rho^2}{I(q)} = \frac{1}{\Phi_H N_H V_H P_H(q)} + \frac{1}{\Phi_D N_D V_D P_D(q)} - \frac{2\chi}{V_0} \qquad (3)$$

$N_i$ and $V_i$ are the number of monomeric units per chain (determined from $M_w$), and monomeric unit volume, for H and D isotopes, respectively. $V_0 = \Phi_H V_H + \Phi_D V_D$ is the average monomer volume. The elevated number of parameters in **eq. (3)** can be reduced significantly by fixing the volumes for both H and D chains according to independent SEC results (**Table 1**):

$$N_i \, V_i = \frac{M_i}{d \, N_A} \tag{4}$$

with $M_i$ the weight-average molecular weight, d the density in g/cm$^3$, and $N_A$ the Avogadro number. The number of parameters can be further limited by coupling the radii of gyration of the two species using their Gaussian statistics and identical monomeric structure:

$$R_{g,D} = R_{g,H} \sqrt{\frac{M_D}{M_H}} \tag{5}$$

There are thus only two free parameters for fits of pure chain melts with **eq. (3)**: $\chi$, and one radius of gyration.

A plausible transition scenario between the two limiting cases – beads and individual chains – is the progressive release of chains from the beads. The average mass of the system of beads and chains thus decreases, which is detectable in the low-q scattering. Furthermore, the polymer chains probably start by interdiffusing on the surface of the beads, which generates a corona of H-chains solvated by D-chains, and vice versa. Such chains thus gain scattering contrast, and latex particles become 'hairy beads'. In this intermediate case, the simultaneous presence of chains and hairy beads leads to the more complex fitting procedure defined in section III.2. **Eq. (3)** is then simplified by using only a single chain contribution:

$$\frac{\Delta\rho^2}{I(q)} = \frac{1}{\Phi_H \Phi_D N V_0 P(q)} - \frac{2\chi}{V_0} \tag{6}$$

This is justified as the molecular weights of the H and D chains are altogether quite close. Note that **eq. (6)** reduces to **eq. (1)** for $\chi = 0$.

The existence of chain scattering, due to released or corona-chains, should result in a high-q power law with a lower exponent than the Porod law, and thus remain visible at high-q. In **Fig. 1**, the evolution of the scattered intensity during annealing is shown for a typical silica-free sample ($A_{62}(0)$, $\Phi_H = 62\%$, see **Table 1** for details). Directly after film formation at 65°C (3 days), the highest intensity is obtained. The decay of the scattering function can be described by a Guinier law, $I = I_o \exp(-q^2 R_g^2/3)$, and we find $R_g = 140$ Å for the typical spatial extension of the objects in the film. Given that the radius of gyration of the precursor latex beads measured independently in aqueous solution is 95 Å, its increase indicates a swelling of the beads by the chains of the surrounding beads, i.e., the beginning of chain interdiffusion. A similar behavior has been reported in the literature [59]. Following **eq. (1)**, the $I_o$ value of 5415 cm$^{-1}$ corresponds to a bead mass quite close to the one of the precursor latex beads (i.e., $\geq 22$ chains out of 23), and in practice no chains have escaped from the beads yet. In the high-q range, however, the expected Porod-law breaks down, and a crossover to a $1/q^2$ regime is found, in close resemblance to scattering of hairy micelles [60, 61]. $R_g$ being of the same order of magnitude as the initial beads, the general structure is still a close-packed assembly of beads with a hairy corona.

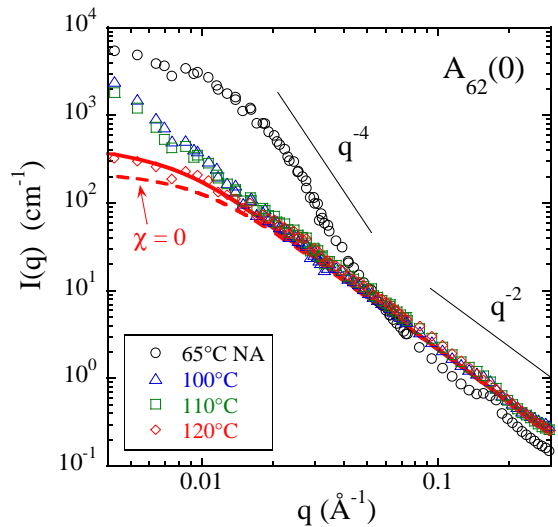

**Figure 1:** Structural evolution of matrix structure during annealing ($A_{62}(0)$, $\Phi_H = 62\%$). After film formation at 65°C (no annealing), then after annealing at 100°C (2 weeks), 110°C (1 week), 120°C (1 week). The solid line is the RPA description, the dotted line represents the theoretical prediction for $\chi = 0$ (Debye).

During annealing at higher temperatures – 100°C (2 weeks), 110°C (1 week), and 120°C (1 week), respectively – the shape of the scattering function evolves considerably in **Fig. 1**. The limiting intensity $I_o$ becomes smaller by more than a factor of ten, indicating vanishing of beads, and the q-range over which chain scattering ($q^{-2}$) is observed widens. Using the Debye

formula – **eq. (2)** – together with the parameter coupling – **eq. (5)** – in **eq. (3)** allows us to fit the scattering function after the strongest annealing in **Fig. 1**. The resulting radii of gyration of the chains in the melt are reported in **Table 2** for the different batches. The $R_g$ values depend on the exact value of $\chi$, the choice of which will be discussed below, but altogether their values are robust within 10%. To check consistency, the values of $R_g$ can be compared to the average mass of the chains. Due to polydispersity in chain mass, the cloud of data points was fitted by a single relationship, $R_g = 1/6 \, C_\infty \, b \, N_{bond}^{1/2}$, where $C_\infty$ is the characteristic ratio, b the bond length (1.53 Å), and $N_{bond}$ the number of bonds [53]. The result is $C_\infty = 12 \pm 3$, which compares favorably with other polymethacrylates, e.g., 9 for PMMA[53]. Using a Kratky-plot, a persistence length can be determined, and a Kuhn length of typically $16 \pm 3$ Å can be deduced (as compared to 17 Å for PMMA[53]). $C_\infty$ deduced from the Kratky-plot method is $9 \pm 2$. Our chain characteristics are thus in the expected range. Note that the precursor solution pH values have been varied from 4 to 9 in the different matrices, in order to agree with those of the nanocomposites discussed later. Given the close range of observed $R_g$ values, they seem to be unaffected by the precursor solution pH.

We now turn to the choice of $\chi$. For an ideal mixture of Gaussian chains ($\chi = 0$), the expected limiting law is shown in **Fig. 1**. For non-zero $\chi$ values, the chain interactions lead to concentration fluctuations, which results in an increased low-q scattering, visible in **Fig. 1**. For very high $\chi$ values, the mixture is no more homogeneous, displaying spinodal decomposition for $\chi \geq \chi_s$. The critical value can be calculated from the point of divergence of I(q=0) in **eq. (3)** assuming identical monomer volumes:

$$\chi_s = \frac{1}{2}\left(\frac{1}{N_H \Phi_H} + \frac{1}{N_D \Phi_D}\right) \quad (7)$$

It follows from **eq. (7)** that low molecular weights minimize the probability for the mixture to undergo phase separation. The $\chi_s$ values are reported in **Table 1**.

The Flory-Huggins parameter can be determined by fitting **eq. (3)** to the SANS-data of samples annealed at sufficiently high temperatures, where chain scattering is predominantly observed. A sound strategy for $\chi$ determination is to perform the same analysis with samples of different H-D-ratio, which is what we have done with the series of samples described in

**Table 1**. In **Fig. 2**, the experimental scattering function of an annealed film (sample $A_{53}(0)$, 120°C, 1 week) is compared to a fit by **eq. (3)**, and good agreement is obtained for $\chi = 5\ 10^{-4}$. The pure form factor ($\chi = 0$) is also shown for this sample. For comparison, the critical value $\chi_s$ is about twice as high ($9.3\ 10^{-4}$).

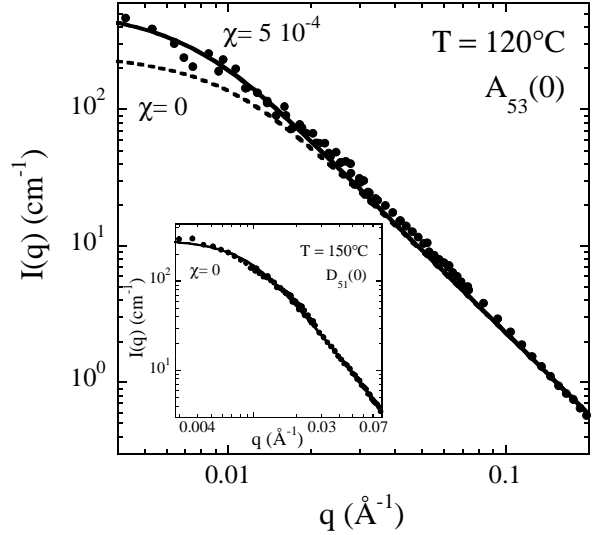

**Figure 2**: Single-chain form factor for the H-D matrix $A_{53}(0)$ after annealing (120°C, 1 week). Line is a fit using the RPA description with $\chi = 5\ 10^{-4}$ (see text for details), dashed line is the theoretical curve in the limit of ideal mixing. Insert: H-D matrix $D_{51}(0)$ after annealing at 150°C, line is a prediction assuming $\chi = 0$.

The same analysis has been performed with the complete series in **Table 1**, and the same $\chi$-value ($5\ 10^{-4}$) was found to reproduce correctly all data, and will be taken from here on as the average monomeric Flory-Huggins interaction parameter of this copolymer at this temperature (120°C). For lower annealing temperatures, only an extrapolation will be proposed. Note that again the pH has no influence on this parameter and that the same radius of gyration is found for the first two samples ($A_{53}(0)$ and $A_{62}(0)$ : same batch, different $\Phi_H$) in **Table 2** with $\chi$ forced to $5\ 10^{-4}$, which illustrates the overall coherence of the method. For completeness, slightly better fits may be obtained by letting $\chi$ evolve freely. Indeed, values between 4.1 and $6.7\ 10^{-4}$ are found, cf. **Table 2**, their average being close to $5\ 10^{-4}$, with an error bar of $10^{-4}$.

| Name | Annealing | χ | $R_g$(H) (Å) | $R_g$(D) (Å) |
|---|---|---|---|---|
| $A_{53}$(0) | 120°C, 1 week | 0.00058 | 170 | 150 |
| | | 0.0005* | 157 | 139 |
| $A_{62}$(0) | 120°C, 1 week | 0.00041 | 146 | 129 |
| | | 0.0005* | 159 | 140 |
| $B_{62}$(0) | 120°C, 2 weeks | 0.00067 | 180 | 180 |
| | | 0.0005* | 166 | 166 |
| $C_{44}$(0) | 120°C, 3 weeks | 0.00063 | 174 | 154 |
| | | 0.0005* | 148 | 131 |
| $D_{51}$(0) | 150°C, 2 weeks | 0.00010 | 186 | 155 |
| | | 0* | 174 | 145 |

**Table 2**: RPA fit parameters for chain melts: interaction parameter χ and radius of gyration for H and D-chains. Value* indicates that this parameter was imposed as an average value compatible with all data sets.

Upon examination of **Table 2**, it becomes clear that longer annealing times at 120°C (up to three weeks, for different $\Phi_H$) have no systematic effect on the interaction parameter. In an independent experiment, a single high-temperature annealing of sample $D_{51}$(0) (150°C, 2 weeks) has been performed, and the result is shown in the inset of **Fig. 2**. The data are seen to be perfectly reproduced by a fit using **eq. (3)** with χ = 0. Note again that an almost imperceptibly better fit is obtained with χ = $10^{-4}$, which is the order of magnitude of the error bar. This suggests that the molecular dispersion obtained after stronger annealing is very close to ideal. Note that this is in qualitative agreement with the generic temperature dependence of χ = A + B/T with B > 0 [53, 55].

**III.2 Modeling of SANS data of pure H-D-matrices**

In this section, it will be shown that the phenomenological evolution of the SANS data can be quantitatively reproduced with an additive model of free chains (RPA) and of an adaptation of the Pedersen model for hairy micelles [60, 61]. The form factor of the latter is mimicked by a sphere with Gaussian chains on the surface. The model allows us to extract detailed information on the fraction and radius of gyration of free chains, as well as on the core-corona structure of the latex beads in the film.

The expression for the form factor of a hairy micelle with non interacting Gaussian corona chains [60, 61] translates into the following normalized function using our notation:

$$P_{hairy}(q) = \frac{1}{V_{HB}^2}\left(V_{core}^2 P_{core}(q) + N_{hair} V_{hair}^2 P_{hair}(q) + 2N_{hair} V_{core} V_{hair} S_{core-hair}(q)\right.$$
$$\left. + N_{hair}(N_{hair}-1)V_{hair}^2 S_{hair-hair}(q)\right) \quad (8)$$

Here $N_{hair}$ is the number of corona chains at the core surface, $V_{hair}$ is the volume of one chain, and $V_{core}$ is the volume of the polymer core. $V_{HB}$ is the (dry) volume of the hairy bead, $V_{HB} = V_{core} + N_{hair}V_{hair}$ due to volume conservation. $P_{hair}(q)$ is the Debye function given in **eq. (2)**, and $P_{core}(q)$ is the form factor of a sphere with radius R

$$P_{core}(q) = 9\frac{(\sin(qR) - qR\cos(qR))^2}{(qR)^6} \quad (9)$$

$S_{core-hair}(q)$ is the interference cross-term between the core and a Gaussian corona chain starting at the surface with a radius of gyration $R_g$*

$$S_{core-hair}(q) = \frac{1-\exp(-q^2 R_g^{*2})}{q^2 R_g^{*2}} \frac{\sin[q(R_g^* + R)]}{q(R_g^* + R)} F_{sphere}(q) \quad (10)$$

where $F_{sphere}(q)$ is the amplitude of the sphere form factor given by the square-root of **eq. (9)**. $S_{hair-hair}(q)$ is the interference between the chains forming the hairy corona

$$S_{hair-hair}(q) = \left(\frac{1-\exp(-q^2 R_g^{*2})}{q^2 R_g^{*2}}\right)^2 \left(\frac{\sin[q(R_g^* + R)]}{q(R_g^* + R)}\right)^2 \quad (11)$$

Due to the contrast situation, only the protruding part of the chains is visible as a hair. In **eqs. (10)** and **(11)**, this absence of (visible) penetration in the core is mimicked by the term $R_g^*+R$.

As we have seen in **Fig. 1**, there is an intermediate state during annealing where the average mass of chains and beads is progressively reduced due to chain escape. Here, this is modeled with a fraction α of free chains in the bulk (out of the bead), whereas the remaining fraction of chains (1-α) is still partially trapped in the latex core. α thus represents the loss of mass of the beads. A second fraction β is needed to describe the partition between core and corona: β = 1 stands for the initial bead with no chains at the surface (all is core), and β = 0 represents a Gaussian star polymer (all is corona). It follows that $V_{core} = \beta (1-\alpha) V_0 = \beta V_{HB}$ and $V_{corona} = N_{hair}V_{hair} = (1-\beta)(1-\alpha)V_0 = (1-\beta) V_{HB}$, where $V_0$ is the volume of the initial latex bead.

To obtain the scattered intensity due to the mixture of hairy beads and free chains, we assume that one can divide our system in two independent space-filling parts: H and D core-corona beads on one side [$I_{hairy}(q)$], and H and D free chains on the other side [$I_{free}(q)$], as illustrated in **Fig. 3**.

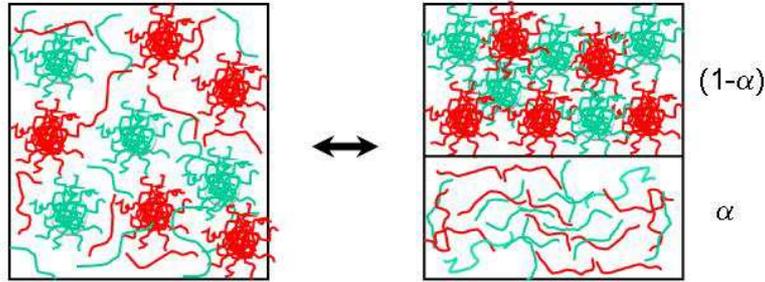

**Figure 3**: Right: schematic representation of the system during annealing. Deuterated components are in red and protonated ones in green. Left: decomposition in two space-filling sub-systems: core-corona beads in proportion (1-α) and free chains in proportion α.

Each case may be described using the formalism presented above. **Eq. (1)** describes the scattering of the subsystem containing the ideal mixture of hairy beads, using the form factor $P_{hairy}(q)$ given in **eq. (8)**. The contrast and the fractions $\Phi_H$ and $\Phi_D$ keep their meaning, only the (dry) volume of the hairy beads, $V_{HB}$, needs to be used for v*. Neglecting the cross-correlation term between the two sub-systems, thus assuming that the structure of chains is not affected by the hairy beads, and vice versa, we add the intensity for each of them weighted by their proportions α (resp. 1-α):

$$I(q) = (1-\alpha)I_{hairy}(q) + \alpha I_{free}(q) \qquad (12a)$$

$$I_{hairy}(q) = \Delta\rho^2 \Phi_D \Phi_H V_{HB} P_{hairy}(q) \qquad (12b)$$

$I_{free}(q)$ of the subsystem of the melt of free chains is given by the RPA-expression, **eq. (6)**. The shape of the resulting signal depends on the form factor of each component, and on α. The total number of parameters seems large at first sight, but again it can be reduced to a few key parameters. Individual bead parameters (mass, density, and radius of H and D beads) are known from independent measurements, as well as chain characteristics (cf. **Table 1**). $\chi$ at 120°C has been fixed to $5\ 10^{-4}$. At lower annealing temperatures, it is higher. In this range, $\chi$ has been fixed to extrapolated values of $7.5 \pm 2\ 10^{-4}$ at 100°C (resp. $6.5 \pm 1.5\ 10^{-4}$ at 110°C), based on the relationship $\chi = A + B/T$, and vanishing $\chi$ at 150°C. Directly after film formation at 65°C, the extrapolated value of $\chi$ exceeds the critical value $\chi_s$, and $\chi$ was set to $\chi_s$. Then the free-chain intensity diverges at $q = 0$, but due to the limited number of free chains ($\alpha \leq ¼$), only little impact on the low-q level of the fit is found. Knowing $\chi$, the parameter $\alpha$ can be read off from $I_0$. The radius of gyration of free chains in the final melt has been reported in **Table 2**, and here their average over H and D chains, $R_{g\_free}$, was imposed (150 Å for batch A), and kept fixed for all stages of annealing. This is consistent with the data, because changing $R_{g\_free}$ directly impacts the high-q prefactor. As for free chains, the radius of gyration of corona chains is coupled to their mass according to **eq. (5)**. Through the conservation of remaining bead volume, the corona chain mass is coupled to the number of corona chains, and to $\beta$. To summarize, besides $\alpha$ which is set independently, the remaining fit parameters are the core fraction $\beta$, and the number of corona chains $N_{hair}$. These two parameters have a specific impact on the medium-q data (curvature, weak oscillations), and can be determined with a relative precision of better than 5%.

| Thermal treatment | Sample | $\alpha$ | $\beta$ | $N_{hair}$ | $R_g$* (Å) (corona) | $R_{core}$ (Å) | $N_{free}$ |
|---|---|---|---|---|---|---|---|
| 65°C, 3 days (NA) | $A_{53}(0)$ | 0 | 0.55 | 90 | 51 | 102 | 0 |
| $\chi = \chi_s$ | $A_{62}(0)$ | 0 | 0.45 | 90 | 56 | 96 | 0 |
| 100°C, 2 weeks | $A_{53}(0)$ | 0.75 | 0.12 | 8 | 120 | 39 | 17 |
| $\chi = 7.5\ 10^{-4}$ | $A_{62}(0)$ | 0.12 | 0.10 | 14 | 171 | 56 | 3 |
| 110°C, 1 week | $A_{53}(0)$ | 0.75 | 0.12 | 8 | 120 | 39 | 17 |
| $\chi = 6.5\ 10^{-4}$ | $A_{62}(0)$ | 0.25 | 0.10 | 10 | 187 | 53 | 6 |
| 120°C, 1 week | $A_{53}(0)$ | 1 | - | - | - | 0 | 23 |
| $\chi = 5\ 10^{-4}$ | $A_{62}(0)$ | 1 | - | - | - | 0 | 23 |

**Table 3**: Hairy bead fit parameters used in **eqs. (12)** for analysis of silica-free matrices. The initial number of chains per bead is $N_0 = 23$. $R_{g\_free}$ was set to its average value after annealing, 150 Å. Only $\alpha$, $\beta$, and $N_{hair}$ are fit parameters, all others are determined by coupling.

The formalism of **eqs. (12)** has been applied to the scattering of the pure H-D-matrices, and the fits are superimposed to the data in **Fig. 4** (solid lines). All the main features, like the low-q limiting intensity, the signal curvature, and the cross-over to $q^{-2}$ chain scattering are found to be reproduced in a satisfactory manner. Note that a simple homogeneous core-shell model would follow a $q^{-4}$ power law in the high-q range, incompatible with the data. Minor differences, like damped oscillations in the intermediate q-range of the model prediction, are due to the use of monodisperse cores and smoothing of experimental data by the spectrometer resolution function. This is shown by comparison to the individual contributions given in eqs.(8-11) to a final scattered intensity, in **Figure 4b**. Depending on the samples, clearly either the Guinier-shoulder of the core form factor, or the oscillations of the same function, induce a minor discrepancy between the data and the model fit.

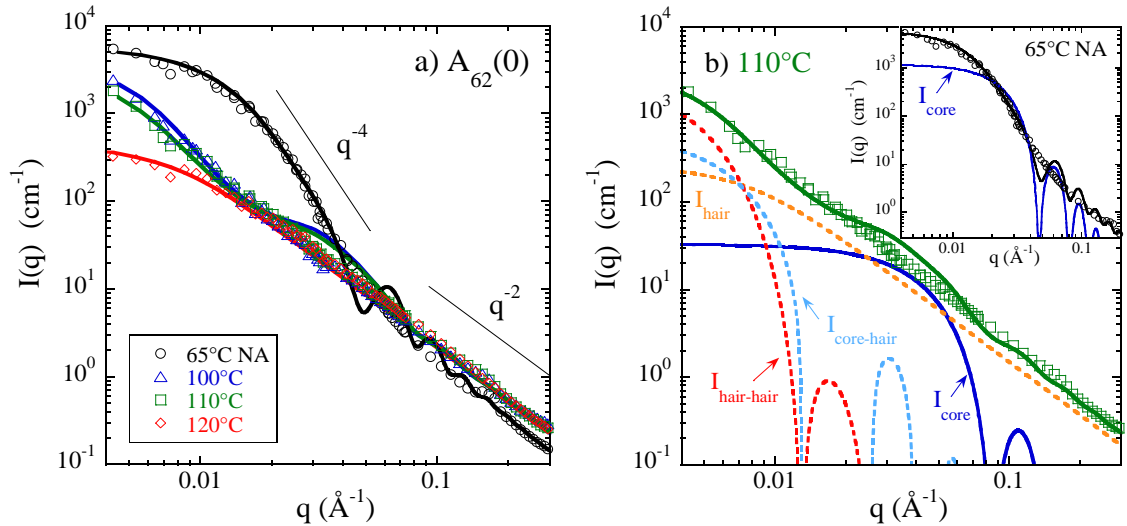

**Figure 4: (a)** Modeling of the structural evolution of the matrix during annealing ($A_{62}(0)$, $\Phi_H = 62\%$). After film formation at 65°C (no annealing), then after annealing at 100°C (2 weeks), 110°C (1 week), 120°C (1 week). **(b)** Decomposition of model fits into different contributions showing the influence of the monodispersity of the core (110°C). Inset: same monodisperse core contribution, without annealing.

The parameters used to fit the data are summarized in **Table 3** for batch A, for two H-fractions. In the course of the annealing procedure, $\alpha$ is found to increase from 0 to 1 at 120°C, where hairy beads have completely disappeared and chains are described by RPA. This is accompanied by the corresponding increase in the number of free chains escaped from each bead. Simultaneously, $\beta$ decreases from an initial value of about ½ to about 1/10. This implies that directly after film formation at 65°C, half of the volume of the latex particles is already 'solvated' by the chains of the surrounding beads. Chain interdiffusion has already started, which is to be expected, given that such latex films possess strong mechanical properties, considerably higher than what would be found for dense assemblies of undeformed colloidal spheres. The number of corona chains, finally, is found to decrease with increasing annealing temperature. Its initial number, of the order of one hundred, is about four to five times higher than the number of chains per bead. This suggests that some chains located at the surface form loops, each being counted as an individual corona chain in the model. As the core radius goes down ($\alpha \rightarrow 0$), and the corona fraction up ($\beta \rightarrow 0$), the overall radius of gyration of the hairy beads is found to increase, as can be seen from the global shape of the curves in **Figure 4a**. Within our model, this swelling of the beads is described by long corona chains (cf. **Table 3**), which are thus more massive, and – by mass conservation – fewer. For one specific sample, their radius of gyration even exceeds the typical $R_g$ of free chains for low-$\alpha$ values. This may be due to stretching (because of the star-like conformation), which is mimicked in our model by more massive chains. It may also be induced by the model which assumes all chains fixed on the core, whereas they may start from anywhere in the corona.

**III.3 Modeling of SANS data describing the polymer matrix structure of silica-latex nanocomposites**

The question of the microstructure of the polymer matrix in nanocomposites can be addressed by SANS [35]. Using appropriate mixtures of H- and D-latex, it is possible to index-match the silica using the method of zero-average contrast [47, 48]. This is why the scattering length densities of the components have been determined with care in section II. The theoretical match point of silica by our latex is $\Phi_H = 51\%$. The small angle neutron scattering of the resulting silica-matched nanocomposites should then originate from the H-D-structure only, and this property has been searched for in the past in order to measure the chain structure in polymer nanocomposites [6-11]. In practice, the issue is more complicated, due to sometimes difficult matching, and to the question of a silica contribution via hole scattering in the matrix, i.e., holes occupied by invisible silica, which may dominate the (mostly low-q) signal at high volume fractions.

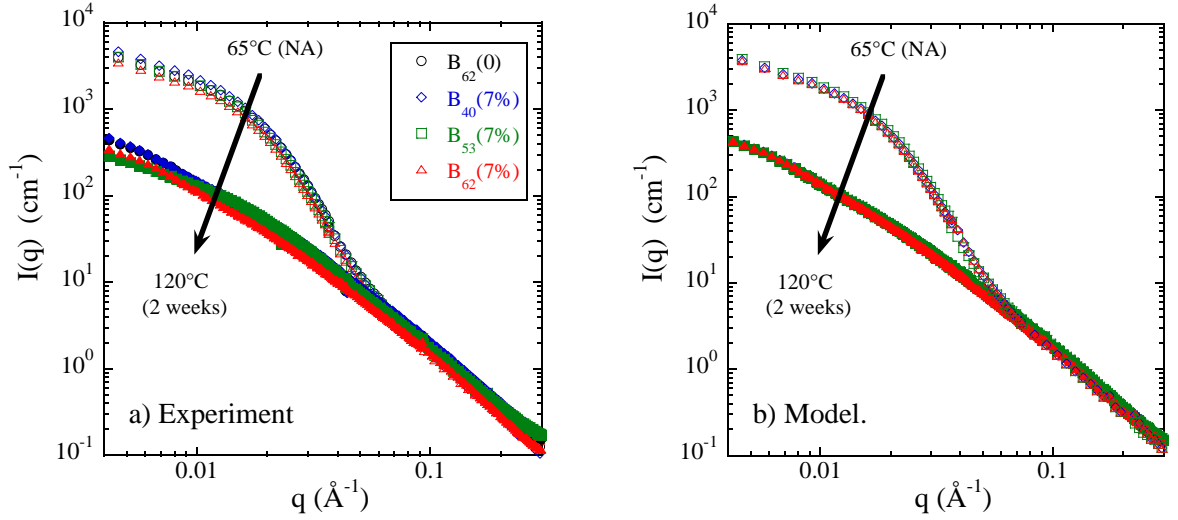

**Figure 5:** (a) Scattered intensities of samples $B_{62}(0)$, $B_{40}(7\%)$, $B_{53}(7\%)$ and $B_{62}(7\%)$ after film formation and after annealing. (b) Theoretical modeling for the same samples following **eq.(13)** showing that H-D polymer scattering dominates the intensity.

Here we start with the analysis of the matching conditions in nanocomposites through measurements around the match point. In **Fig. 5a**, the scattering of a pure H-D-matrix (sample $B_{62}(0)$) is compared to nanocomposites at $\Phi_{si} = 7\%$ : $B_{40}(7\%)$, $B_{53}(7\%)$ and $B_{62}(7\%)$, before (film formation at 65°C) and after annealing (120°C, 2 weeks). The scattering for different $\Phi_H$ is seen to remain very close, compared to the strong variation induced by the annealing procedure. To emphasize the satisfying matching conditions, we have calculated the theoretical intensities expected for such samples, based on rescaled additive contributions of the silica and the matrix:

$$I_{theo}(q) = \left(\frac{\Delta\rho_2}{\Delta\rho_1}\right)^2 I_{si}(q) + \frac{\Phi_H{'}\Phi_D{'}}{\Phi_H\Phi_D}(1-\Phi_{si})I_{matrix}(q) \qquad (13)$$

where $I_{si}(q)$ has been measured in a pure H-matrix with pH 4 (cf. scattering curve in appendix), under contrast condition $\Delta\rho_1 = \rho_{si} - \rho_H$, at $\Phi_{si} = 7\%$, rescaled to contrast $\Delta\rho_2 = \rho_{si} - \rho_{HD}$; $I_{matrix}(q)$ is the scattering of the silica-free matrix measured at $\Phi_H = 62\%$, rescaled to $\Phi'_H$ (i.e., 40%, 53%, 62%). The predictions in **Fig. 5b** are also very close for H-fractions between $\Phi_H = 40\%$ and 62%, which illustrates that matching is satisfying over this range. We have checked that higher deviations ($\Phi_H = 70\%$ and above) distort the intensity curves considerably, due to the then visible silica contribution. The contribution of the matched silica can thus be considered negligible, and we can apply the analysis developed in section III.2 to the scattering of the remaining matrix fraction of the samples by dividing experimental intensities by $(1-\Phi_{si})$.

The evolution of the polymer structure with annealing of silica-containing samples with various $\Phi_H$ is shown in **Figure 6,** together with model fits. The data either directly after film formation at 65°C, or after annealing of one or two weeks at 120°C are plotted. The temperature was chosen in the light of the preceding section. In **Fig. 6**, the scattering curves of nanocomposites show the same evolution as the pure matrices (cf. **Figs. 1** and **5**): the low-q intensity decreases, and chain scattering progressively dominates the signal. Note that the matrix made with exactly the same batch, $B_{62}(0)$, is shown in **Figure 5**.

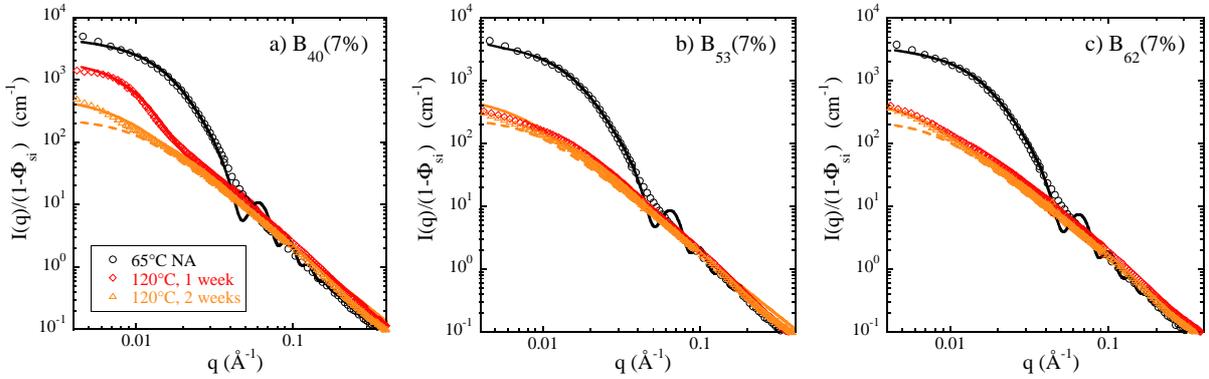

**Figure 6**: Structural evolution of polymer structure in nanocomposites with $\Phi_{si}$ = 7% after film formation and annealing at 120°C for one and two weeks. **(a)** $B_{40}(7\%)$, **(b)** $B_{53}(7\%)$, **(c)** $B_{62}(7\%)$. Solid lines are theoretical models explained in the text. The dotted line represents the theoretical prediction for $\chi = 0$ (Debye).

The fitting procedure of nanocomposites using **eqs. (12)** is the same as in section III.2, and fits are superimposed to the data in **Fig. 6**. The fit parameters for all samples are given in **Table 4,** together with those of the corresponding silica-free matrix. The interaction parameter $\chi$ was set, as before, to $\chi_s$ after film formation and to $5\ 10^{-4}$ at 120°C. The radius of gyration of the free chains in coexistence with the latex beads has been determined from the longest annealing (120°C, 2 weeks) using **eq. (6)** for each nanocomposite. These values are compatible (within 15%) with the value of the pure matrix, and have been imposed to weaker annealing (cf. caption of **Table 4**). The decrease in average mass is described by an increase in $\alpha$ from about ¼ to 1 after one or two weeks at 120°, which implies the complete disappearance of the latex core. After one week, some samples (the pure matrix $B_{62}(0)$ and $B_{40}(7\%)$) still display a scattering function intermediate between beads and chains, which corresponds to $\alpha \approx ½$. The core-corona parameter $\beta$ – which quantifies the remaining core fraction and thus the degree of chain interdiffusion – decreases from about ½ to 1/6 for these two samples. The number of corona chains is equivalent to the one of the silica-free matrices discussed in section III.1: it decreases from about 85 to about 30. In parallel, the radius of gyration of the corona chains grows from about 50 to 100 Å. For all samples, after disappearance of the core (annealing for 2 weeks), the signal evolves to free chain RPA-behaviour (**eq. (3)**). The number of free chains having escaped from the beads is typically 5 directly after film formation, and reaches the maximum number $N_0 = 23$ after annealing.

| Thermal treatment | Sample | α | β | $N_{hair}$ | $R_g^*$ (Å) (corona) | $R_{core}$ (Å) | $N_{free}$ |
|---|---|---|---|---|---|---|---|
| 65°C no annealing $\chi = \chi_s$ | $B_{62}(0)$ | 0.24 | 0.57 | 70 | 54 | 95 | 6 |
| | $B_{40}(7\%)$ | 0.15 | 0.55 | 75 | 52 | 97 | 3 |
| | $B_{53}(7\%)$ | 0.22 | 0.50 | 90 | 54 | 91 | 5 |
| | $B_{62}(7\%)$ | 0.27 | 0.52 | 100 | 52 | 91 | 6 |
| 120°C, 1 week $\chi = 5\ 10^{-4}$ | $B_{62}(0)$ | 0.45 | 0.15 | 28 | 103 | 54 | 10 |
| | $B_{40}(7\%)$ | 0.43 | 0.10 | 30 | 95 | 48 | 10 |
| | $B_{53}(7\%)$ | 1 | - | - | - | 0 | 23 |
| | $B_{62}(7\%)$ | 1 | - | - | - | 0 | 23 |
| 120°C, 2 weeks $\chi = 5\ 10^{-4}$ | $B_{62}(0)$ | 1 | - | - | - | 0 | 23 |
| | $B_{40}(7\%)$ | 1 | - | - | - | 0 | 23 |
| | $B_{53}(7\%)$ | 1 | - | - | - | 0 | 23 |
| | $B_{62}(7\%)$ | 1 | - | - | - | 0 | 23 |

**Table 4**: Hairy bead fit parameters used in **eqs. (12)** for the analysis of H-D-nanocomposites containing $\Phi_{si}$ = 7%. The initial number of chains per bead is $N_0 = 23$. The radius of gyration of free chains is set to the average value determined for each sample after annealing: $R_{g\_free}$ = 166 Å, 152 Å, 170 Å and 183 Å for matrices $B_{62}(0)$, $B_{40}(7\%)$, $B_{53}(7\%)$, $B_{62}(7\%)$, respectively. Only α, β, and $N_{hair}$ are fit parameters, all others are determined by coupling.

In a previous article, we have investigated the film structure of blends of H- and D-latex beads which demix during annealing. The demixing kinetics was found to proceed during annealing for nanocomposites with low amounts of silica ($\Phi_{si}$ = 5%), and to be blocked in presence of higher amounts $\Phi_{si}$ = 15% [34]. For the present system of compatible latex beads, we have performed an analogous study with a high silica fraction ($\Phi_{si}$ = 20%). Note that the samples, $A_{53}(20\%)$ and $A_{62}(20\%)$, were designed to have a silica structure comparable to the one of the 7% samples by varying the pH. This was checked by SANS with the purely hydrogenated samples, cf. data reported in the appendix. The evolution of the polymer structure of $A_{53}(20\%)$ and $A_{62}(20\%)$ in the course of the annealing procedure – 65°C NA, 100°C (2 weeks), 110°C (1 week), and 120°C (1 week and 2 weeks) – is shown in **Fig. 7**.

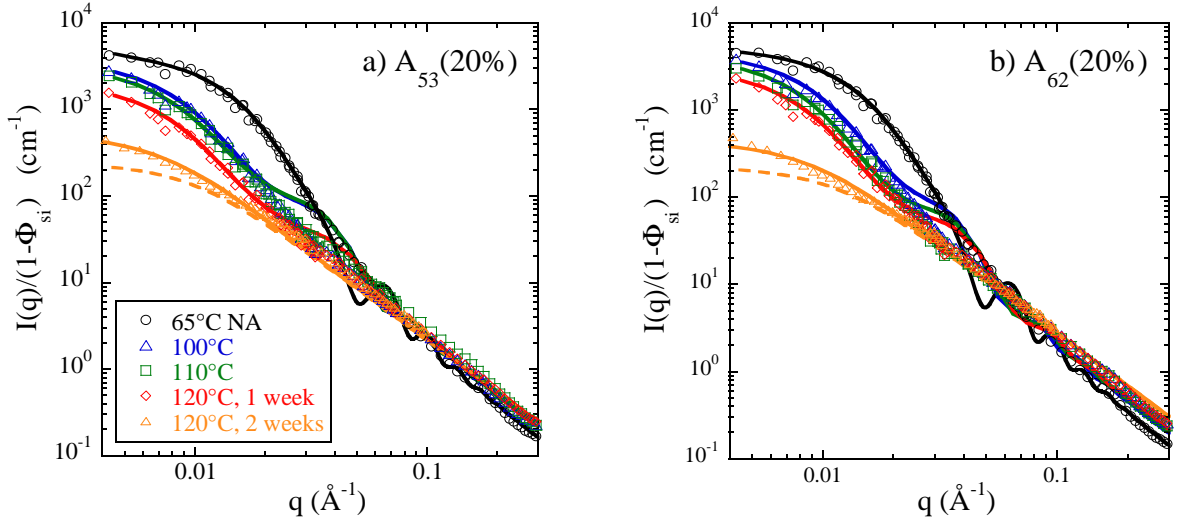

**Figure 7:** Structural evolution of polymer structure in nanocomposites with $\Phi_{si}$ = 20% after film formation and annealing at 100°C (2 weeks), 110°C (1 week), 120°C (1 week) and 120°C (2 weeks). **(a)** $A_{53}$(20%), **(b)** $A_{62}$(20%). Solid lines are theoretical models as explained in the text. The dotted line represents the theoretical prediction for $\chi = 0$ (Debye). The corresponding matrix $A_{62}$(0) is shown in **Fig. 1**.

The scattering curves displayed in **Figure 7** show a delayed evolution with respect to the silica-free matrices (or 7%-nanocomposites) during annealing. The initial structure directly after film formation (NA) resembles to the one previously encountered for hairy beads (cf. **Figs. 1** and **6**). During annealing of the nanocomposites at higher T, however, the low-q intensity decreases only slowly, and longer annealing at 120°C is needed to achieve a RPA-behavior ($\alpha$=1). The resulting radius of gyration of the free chains are $R_g$(H) = 163 Å (resp. 140 Å), and $R_g$(D) = 144 Å (resp. 124 Å) for $A_{53}$(20%) and $A_{62}$(20%) respectively, using the interaction parameter $\chi$ (fixed to 5 10$^{-4}$).

We have again reproduced the scattered intensities using the formalism of **eqs. (12)**, based on the average (over H and D) radius of gyration of the free chains, and the fit parameters are summarized in **Table 5**. The NA-curve can be described by similar parameters as in **Tables 3** and **4**: a small $\alpha$ in the range 0.05 - 0.15, the core-corona parameter $\beta$ about ½ (as before in all cases), and 85 chains or loops in the corona, of typical radius of gyration $R_g$* ≈ 55 Å. In **Fig. 7**, the moderate decrease in average mass is reflected by a moderate increase in $\alpha$, up to 120°C during one week. Contrary to the silica-poor nanocomposites, the core is thus not completely destroyed under these conditions, and as a consequence, the number of free chains having escaped from a bead is limited to 10, out of $N_0$ = 23 making up the initial bead. Only after two weeks at 120°C, the core disappears and chain scattering is recovered. In this context, it may be noted that the RPA-fit of the chain scattering in presence of silica is empirical, as it implicitly relies on translational invariance of the system[62], whereas some heterogeneities in chain mobility probably exist (cf. section III.4).

| Thermal treatment | Sample | α | β | $N_{chain}$ (corona) | $R_g^*$ (Å) (corona) | $R_{core}$ (Å) | $N_{free}$ |
|---|---|---|---|---|---|---|---|
| 65°C, 3 days (NA) $\chi = \chi_s$ | $A_{53}$(20%) | 0.14 | 0.46 | 85 | 53 | 92 | 3 |
| | $A_{62}$(20%) | 0.05 | 0.47 | 85 | 55 | 96 | 1 |
| 100°C, 2 weeks $\chi = 7.5\ 10^{-4}$ | $A_{53}$(20%) | 0.25 | 0.23 | 28 | 103 | 70 | 6 |
| | $A_{62}$(20%) | 0.10 | 0.24 | 40 | 94 | 75 | 2 |
| 110°C, 1 week $\chi = 6.5\ 10^{-4}$ | $A_{53}$(20%) | 0.29 | 0.24 | 22 | 113 | 69 | 7 |
| | $A_{62}$(20%) | 0.15 | 0.19 | 30 | 109 | 68 | 3 |
| 120°C, 1 week $\chi = 5\ 10^{-4}$ | $A_{53}$(20%) | 0.45 | 0.15 | 18 | 116 | 54 | 10 |
| | $A_{62}$(20%) | 0.28 | 0.18 | 25 | 111 | 63 | 6 |
| 120°C, 2 weeks $\chi = 5\ 10^{-4}$ | $A_{53}$(20%) | 1 | - | - | - | 0 | 23 |
| | $A_{62}$(20%) | 1 | - | - | - | 0 | 23 |

**Table 5**: Hairy bead fit parameters used in **eqs. (12)** for the analysis of H-D-nanocomposites containing $\Phi_{si}$ = 20%. The initial number of chains per bead is $N_0 = 23$. The radius of gyration of free chains is set to the average value determined for each sample after annealing: $R_{g\_free}$ = 154 Å and 134 Å for $A_{53}$(20%) and $A_{62}$(20%), respectively. Only α, β, and $N_{hair}$ are fit parameters, all others are determined by coupling.

Simultaneously, up to annealing of one week at 120°C, the core-corona parameter β decreases to 1/6. This shows that interdiffusion of corona chains with chains from neighbouring beads proceeds approximately in the same manner as in silica-poor samples, but does not reach the complete disappearance of the core. The number of corona chains decreases from 85 to 20, i.e., again less than with silica-poor samples. In parallel, the radius of gyration of the corona chains, $R_g^*$, increases to 100 - 120 Å, i.e., to about 80% of the spatial extent of free matrix chains, as in the case of nanocomposites with 7% of silica for similar α values. The behaviour seems to be different in the case of matrices, where $R_g^*$ increases above the radius of gyration of the free chains for α ≈ 0.1 – 0.2. This could be the signature of the impact of silica on chain interdiffusion.

To summarize, for $\Phi_{si}$ = 20%, we have found radii of gyration compatible with the pure matrix conformations, as well as with the 7%-silica nanocomposites, within error bars.

**III.4 Chain interdiffusion, ~~delayed dynamics~~ and conformation**

It has been shown in the preceding section that the evolution of the structure with annealing at increasing temperatures (65° up to 120°C) is slower in the case with strong silica loading (20%v) as opposed to 7% or silica-free samples. In all cases, the intermediate structures, which reflect the distribution of deuterated and hydrogenated polymer in the samples, could be modeled by a formalism – **eqs. (12)** – based on the Pedersen model for hairy beads, and RPA for the free chains. Two robust results of the fitting procedure are $\alpha$ and $\beta$. Their value can be translated directly into the core volume of the latex particles $V_{core}$, or the corresponding radius $R_{core}$. Decrease of the latter illustrates the progressive interdiffusion of the latex beads. In **Tables 3** and **4**, the radius of the latex core can go to zero after one week at 120°C for matrices and 7%-nanocomposites, whereas it never reaches zero for 20%-nanocomposites (**Table 5**) for the same annealing. In this case, the core vanishes only after two weeks. Observations of similar delay has been made in our previous study on an incompatible latex system, where the kinetics of phase-separating latex zones was investigated [34]. At high silica-content, zone-growth was impeded, but it was unclear if this could be traced back to arrested dynamics at the local scale (chains), or of larger zones (beads). Here, individual chain dispersion is quickly reached in pure matrices and low-silica nanocomposites, suggesting that it is interdiffusion of individual chain ~~dynamics~~ which is delayed at 20%.

This observation of delayed chain ~~dynamics~~ interdiffusion in presence of silica may be explained by several mechanisms, and there is some controversy in the literature. One can imagine, e.g., a sticky interaction of polymer chains on the silica nanoparticles, which would reduce chain mobility.[63, 64] These authors argue that chain adsorption restricts, e.g., reptational dynamics, and thus also polymer flow around hard particles. In some sense, a zone of restricted flow may develop across the sample. Alternatively, a 'bottleneck' explanation has been put forward [65], where the three-dimensional silica structure is thought to constrain possible motion of the polymer latex molecules. This may apply to the a priori percolated silica structure at 20%. Our observations may also be explained with NMR investigations of nanocomposite model systems developed over the past decade [24-29], which postulate another type of zone of restricted dynamics. These authors propose the existence of a frozen polymer layer close to the filler surface, of estimated nanometric thickness (typically a few nm) at temperatures not too far above the bulk glass-transition temperature. In our case, samples are well above the bulk $T_g$ during annealing, and the frozen layer thickness should be smaller. This layer leads to a higher volume fraction of hard matter, and thus can immobilize substantial parts of the nanocomposite sample on our time scale of observation. Exact values depend on the dispersion of the silica in the sample, as well as on the temperature-dependent layer thickness [24]. If the data shown in this article is not suitable for deciding in favour any of the cited explanations, the existence of a delayed chain interdiffusion itself is unambiguously proven. ~~In any event, nanometric glassy layers seem to be a plausible explanation for the observed slowed down dynamics of chains.~~

In the context of nanocomposite reinforcement, the conformation of macromolecules in presence of fillers has been discussed in the literature [9, 10, 66], but no generally accepted trend has been found. An overview has been recently given by Nusser et al.[11] who considered both simulations and experimental aspects based on scattering experiments. They emphasized that the polymer-filler size ratio is a key parameter influencing the polymer structure. Here, we are concerned with filler particles of size comparable to the polymer chains and we will restrict the discussion of (controversial) literature results to this regime. On one hand, a decrease of the chain dimensions with respect to the unfilled polymer was observed by Nakatani[6] and Nusser[11] by means of SANS measurements. In the first case, the authors[6] used a data treatment based on the high concentration method to extract the chain radius of gyration in PDMS/polysilicate fillers blend. Note that this is a specific case because the filler is liquid at room temperature. In the second one, Nusser et al.[11] obtained the chain conformation in silica/PEP nanocomposites by the use of ZAC method. In this system, the $R_g$ decrease was weak for silica loading: 2% reduction at $\Phi_{Si} = 18\%$v. On the other hand, various experiments using the ZAC method report that, in this regime, the chain size is not affected in the presence of silica (Sen[8], Jouault[7]). In agreement with these findings, our results seem to indicate that there is not a strong enough tendency in the evolution of $R_g$ with filler content in order to be detectable. Changing $R_g$ has a direct influence on the high-q scattering, which cannot be compensated by any other parameter in our description. We have thus estimated our error bar on $R_g$ of the free chains to ±15% using a RPA fit. For the nanocomposites with 7% silica (batch B), the $R_g$ of the free chains stays comparable to the one of the matrix, i.e., 166 Å for $B_{62}(0)$. At 20% silica content (batch A), the average (over H and D) $R_g$ of free chains is $R_g$ = 154 Å (resp. 134 Å) for $A_{53}$ (resp. $A_{62}$). These values can be compared to the value of the corresponding matrices (Table 2), which is about 150 Å. As mentioned before, the chain conformation thus stays within the same bounds of ±15% for the radius of gyration, i.e., chain conformation seems to be independent of silica concentration. Jouault et al.[7] draw attention to the role of the filler structure. We point out that our system has a similar aggregation state as their 5%v-silica/PS nanocomposites ($N_{agg} \approx 10$ versus $N_{agg} \approx 7$ in our case, see appendix), and the same unchanged $R_g$ is observed. To summarize, using appropriate contrast-matching in a system where the scattering data are not spoiled by additional terms [8, 11], it is possible to follow the chain radius of gyration even in presence of considerable amounts of silica filler.

## IV. Conclusion and perspectives

The evolving structure during annealing of silica-latex nanocomposites has been studied by SANS under zero-average contrast conditions for the silica nanoparticles. A quantitative structural model based on additive contributions of core-corona ('hairy') latex beads and of free chains in a melt has been developed. The model is demonstrated to nicely reproduce the data over the whole q-range and in absolute units, with a reduced number of free parameters describing the structure of the shrinking latex beads. Our analysis of silica-free matrices and nanocomposites with low (7%) and high (20%) silica volume fractions shows that chain interdiffusion proceeds until complete disappearance of the core, with a delay in presence of 20% of silica. ~~Nevertheless, at 20%, the chain dynamics appears to be delayed.~~

For nanocomposites, evidencing by SANS that the silica filler influences the chain dynamics is interesting for the molecular understanding of the reinforcement effect. Another relevant aspect resides in the still open question of chain conformation in hard filler environments. In this article, it has been shown that annealing at high temperature for sufficiently long times pushes the system towards individual chain dispersion. We have been able to measure chain conformation at different silica volume fractions in a given state of aggregation (typically ten primary silica particles per aggregate, percolating at high concentration). In particular, it is found that the radius of gyration is unaffected by the silica loading under these conditions. It is now hoped that our system will allow obtaining insight in the structure of molecularly dispersed chains in nanocomposites, as a function of dispersion or aggregation, and chain mass.

**Acknowledgements:** Stock solutions of silica were a gift from Akzo Nobel. Beamtime was accorded at Laboratoire Léon Brillouin (Saclay) and Institut Laue Langevin (Grenoble). This work was conducted within the scientific program of the European Network of Excellence Softcomp: 'Soft Matter Composites: an approach to nanoscale functional materials', supported by the European Commission. MT thanks ILL for financing her PhD within the ILL international PhD-program. Financial support by a "chercheur d'avenir" grant (JO) of the Languedoc-Roussillon region is gratefully acknowledged. The authors are also very grateful to Jacques Jestin (LLB, Saclay) who performed the final measurements on the nanocomposites with high silica loading.

**APPENDIX:**

**A. Contrast variation**

The scattering length density of both H- and D-latex was determined by external contrast variation in $H_2O/D_2O$ mixtures (**Figure A1**).

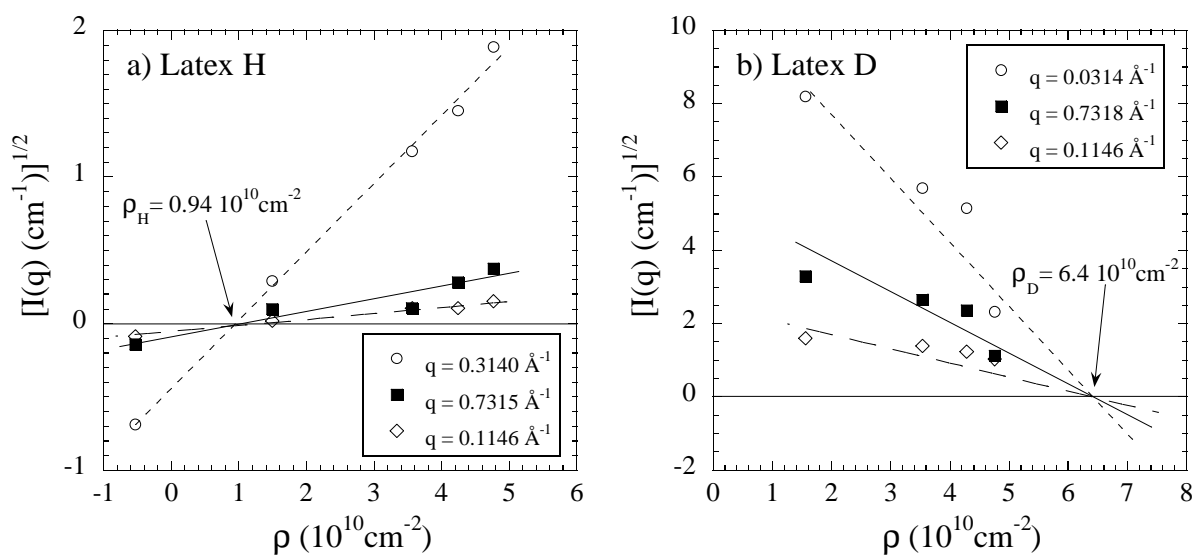

**Figure A1**: SANS contrast variation study of (a) protonated and (b) deuterated latex in a mixture of $H_2O/D_2O$.

## B. Silica structure at 7% and 20%

The structure of the silica filler at 7% and 20% was measured using SANS with a hydrogenated latex matrix. Following the general shape of the aggregation diagram [34], the aggregation is expected to be similar at low $\Phi_{si}$ and low pH, and at high $\Phi_{si}$ and high pH. The data, in a normalized presentation $I(q)/\Phi_{si}$, are shown in **Figure B1**.

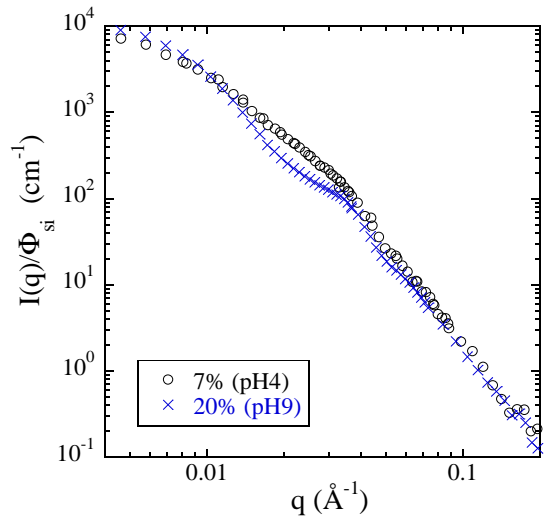

**Figure B1**: Normalized scattered intensity of silica-latex nanocomposites (7% with pH4, and 20% with pH9, both H-matrices) as a function of wavevector q.

As can be concluded from the overlap of large parts of the scattered intensities, the structure is indeed rather similar. A detailed analysis, however, is difficult, due to the presence of unknown, and rather featureless (no peak) structure factors in both cases. The presence of a correlation hole at intermediate q with 20% silica suggests that silica beads are aggregated, and the similar Guinier domain may indicate comparable sizes. In the case of the lower $\Phi_{si}$, it corresponds to average aggregates ($R_g$ = 180 Å) of about seven nanoparticles, i.e., having a compacity of about 32%. At the higher volume fraction of 20%, these aggregates come necessarily into contact, and form a network structure. To summarize, these aggregates do not change much upon simultaneous increase in concentration and decrease in pH, but approach and percolate. On the length scale probed in SANS, they have thus a comparable primary structure which is more or less diluted.